Études coordonnées par
**Jacques SPINDLER**
avec la participation de
**David HURON**

# L'ÉVALUATION
# DE L'ÉVÉNEMENTIEL TOURISTIQUE

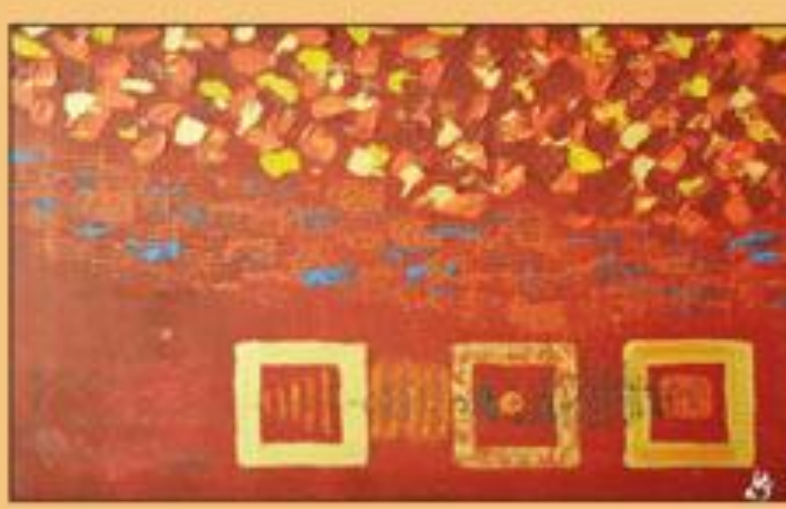

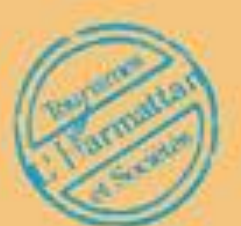

# Chapitre 3

# Barcelone face à la globalisation, comment penser la ville par l'organisation et l'évaluation des grands événements ?

**De l'instrumentalisation des Jeux olympiques et des expositions internationales À la recherche des temps d'approche – temporalités pour l'évaluation d'un méga-événement**

**Par Patrice BALLESTER[1]**

*«Il serait très fâcheux, Messieurs, que les dépenses souvent exagérées faites dans les dernières Olympiades et dont une bonne partie d'ailleurs représente l'Édification de monuments permanents inutiles - le provisoire suffirait pleinement et la seule conséquence est d'inciter ensuite pour utiliser les monuments à multiplier les occasions d'y grouper les foules - il serait, dis-je, très fâcheux que ces dépenses détournassent des (petits) pays de se mettre sur les rangs pour accueillir les Jeux olympiques de l'avenir ».*

Pierre de Coubertin, *Revue Olympique*, avril 1911, pp.59-62.
Repris par la Commission d'étude des Jeux olympiques.
Rapport à la 115ᵉ Session du CIO Prague, juillet 2003.

C'est avec la lecture d'un article de presse, une courte dépêche du journal du Figaro.fr, trois jours avant la victoire de Londres face à Paris lors de la réunion du CIO à Singapour pour l'organisation des Jeux olympiques de 2012, que l'on s'aperçoit des raccourcis inhérents à la mondialisation et de la perception de ces très grands projets d'aménagement et de société que sont les olympiades : *« Barcelone sourit, Atlanta attend, Barcelone, hôte des Jeux en 1992, peut dire merci aux olympiades. C'est la plus grande gagnante des villes organisatrices sur les trente dernières années »*[2]. Durant les derniers mois avant l'échec de la candidature de « Paris 2012 - L'Amour des Jeux », la presse sportive spécialisée s'est penchée sur les effets de tels événements concernant les conséquences sociales et culturelles. Ce titre accrocheur nous permet de mieux cerner ce que la société civile et les politiques par l'intermédiaire des médias retiennent et surtout mettent en avant pour créer une représentation positive de l'après événement en quelques lignes. En récession, comme le reste de l'Espagne, après la chute du régime de Francisco Franco et la transition démocratique, la capitale catalane retrouve la croissance quand elle obtient les Jeux. Les investissements sont considérables. Le taux d'emploi explose avec 21,4 % entre 1986 et 1991. Pendant les olympiades de Barcelone, le taux de chômage est au minimum avec un taux de 8,4 % de la population active en quête de travail. Certes, l'économie espagnole ralentit après les Jeux, mais le tourisme devient une activité très importante pour la ville. Barcelone est remodelée, embellie et sa capacité hôtelière a crûe de 30 % pour l'événement renforçant le tourisme de loisirs et d'affaires. Ce qui lui permet d'accueillir les 2,7 millions de visiteurs recensés en 1995 contre à peine 1,750 millions en 1991 et de créer une toute nouvelle agence de tourisme moderne et innovante ciblant de nouvelles clientèles, tout en misant sur le tourisme de masse et urbain.

---

[1] Patrice Ballester est doctorant au laboratoire GEODE UMR 5602 MSH, Université Toulouse Le-Mirail, professeur d'histoire-géographie dans le secondaire et chargé d'enseignements en géographie.
[2] Le Figaro, [04 juillet 2005], Rédaction.

Pour Barcelone 1992, les olympiades sont utilisées pour promouvoir l'excellence de projets urbanistico-touristiques et immobiliers de toutes sortes, y compris la création d'espaces publics emblématiques attractifs comme son nouveau front de mer. Nous assistons à une instrumentalisation d'un méga-événement à des fins dépassant la simple question du sport et des retombées culturelles. À ce titre, quatre ans plus tard, en 1996, Atlanta accueille à son tour les Jeux olympiques. Des espérances démesurées sont annoncées, 4 milliards de retombées économiques directes et indirectes sont envisagées, mais aussi plus de 100 000 emplois nets. Mais dans la réalité, les retombées sont surestimées, et cela est souvent le cas concernant les olympiades, chiffrages et statistiques sont aléatoires, voire contradictoires parfois, suivant l'institut ou la société d'audit réalisant une évaluation.

Soit deux villes, deux évaluations et deux destinées géoéconomiques. Une appréciation est faite somme toute juste de la situation de ces deux métropoles, ceci 15 ans après les JO de Barcelone en 1992 et 9 ans après ceux d'Atlanta. Les expositions internationales ou universelles et les Jeux olympiques sont devenus des moyens de faire et de penser la ville par l'organisation de grands événements touristiques[3]. Ils attirent des touristes du monde entier, tout en polarisant l'attention du monde et de la scène médiatique pendant un temps non négligeable, cette stratégie de marketing urbain étant devenu de nos jours un moyen d'action très attendu par tous les réseaux sociaux et décideurs comme Tour-Opérateurs et professionnels du tourisme.

L'événement interroge les hommes qu'il soit politique, culturel ou touristique. Il possède une signification propre en tant qu'il commence quelque chose tout en manifestant une volonté, une nouvelle possibilité de créer, de rencontrer et de surprendre. L'événement est en faîte un « *avènement qui atteint tout* » (De Certeau, l969)[4] s'intégrant généralement dans de long processus ou phase de l'évolution des sociétés sur le plan de sa structure sociétale. Or, si l'événement existe, il est significatif donc évaluable.

Comment évaluer des Jeux olympiques ou une exposition internationale ? Peut-on évaluer l'avant (à partir de la prise de position, dossier de candidature), le pendant (le déroulement de la manifestation touristique, culturelle ou sportive) et l'après (la clôture de la fête et ses conséquences directes et indirectes sur la société) ? Comment alors évaluer l'ensemble des dimensions de l'événementiel reposant généralement sur de grandes opérations touristiques et urbanistiques ? Il faut se demander ce que l'on entend dans un premier temps par les termes d'*évaluation* et d'*événementiel* touristique, tout en mettant en avant le terme de *prétexte* comme une sorte de recherche induite par les acteurs du projet. Dans un second temps, quels ont été les différentes configurations et modèles d'évaluation et de calcul concernant ce type d'événement majeur depuis une dizaine d'années, notamment par les bureaux d'audit internationaux ? Que pouvons-nous en retirer et proposer comme grille d'analyse ? Enfin, dans un dernier temps, nous prendrons comme terrain d'étude une ville – capitale – port de commerce devenue une sorte d'archétype dans le cadre d'organisation de méga-événement à vocation touristique que sont les Expositions universelles ou internationales et Jeux olympiques, à savoir Barcelone et trois dates très importantes pour la capitale catalane : 1888, exposition universelle, 1929 expositions universelle et internationale, 1992 les Jeux olympiques modernes et 2004 le Forum universel des cultures de l'Unesco, exposition internationale[4].

---

[3] Masboungi Ariella, « Comment penser la ville par l'événement ? », Paris, 2004, Le moniteur. [4] De Certeau Michel, *La prise de parole*, Paris, Seuil, 1969
[4] Nous ne prendrons pas en compte les olympiades populaires de Barcelone en 1936 et les Jeux Méditerranéens de 1955.

Grâce à ces différentes étapes, nous répondrons à cette question que l'on peut qualifier parfois d'ordre philosophique : quels sont les effets durables de l'éphémère ? Qu'est-ce qu'un événement réussi ? En quoi ont consisté les jeux des acteurs globaux (réseaux) à travers la volonté de bien faire et surtout d'optimiser un budget, tout en produisant des bénéfices de toutes sortes à la fois matériels, immatériels, quantifiables et qualifiables ?

Entre les attentes, les valeurs et les représentations que l'on se fait de « son » événement et la réalité de sa tenue, la ville de Barcelone porte notre attention comme un exemple à retenir et à analyser dans ses conséquences et parfois comme recours par d'autres métropoles mondiales depuis 1992, rappelant à leur tour dans leurs dossiers officiels de candidature la réussite de ces olympiades méditerranéennes à des fins de justification.

## 1. L'événementiel touristique et son évaluation : une question de sémantique et de temporalité

Trois termes sont à définir par leur complexité, mais aussi à associer dans le temps, l'espace et les institutions où elles sont employées : un événement, une évaluation, un prétexte.

### *L'événement ?*

L'événement est tout simplement ce qui arrive, ce qui advient à une date et en un lieu précis. L'événement peut être considéré comme un mode d'individualisation de l'histoire de l'Homme qui se situe directement face à un *« event »*, un fait, une action segmentant la continuité du temps historique et inaugurant un changement dans l'ordre des choses[5]. Si les sciences sociales se sont longtemps méfiées de celui-ci, il est manifeste de constater qu'il existe un retour de l'événement et de l'événementiel en relation avec le retour de l'acteur[6]. À cela, avec *« l'avènement des loisirs »* (Corbin, 1989) et celle de la société de consommation reposant sur *« l'American way of life »*, des projets d'aménagement s'intégrant dans les mécanismes de la mondialisation[7] sont portés par les dirigeants de grandes métropoles européennes et occidentales pour redynamiser leur trame urbaine et requalifier une image touristique de leur ville, tout en donnant à leurs concitoyens des infrastructures de qualité. Le méga-événement est invoqué, convoqué, voire parfois imploré par certains décideurs politiques comme une « idée » à concrétiser.

Les acteurs politiques, sportifs et culturels utilisent l'évaluation ante-événement ou post-événement concernant d'autres manifestations géantes de même type, ayant eu lieu dans une autre métropole plus ou moins comparable dans ses fonctions, formes et poids démographique et politique. Ils intègrent très souvent des comparaisons proches dans le temps avec les anciennes tenues de manifestations ou bien très éloignées (parfois plus d'un siècle) à des fins de manipulations.

---

[5] Dictionnaire de géographie et de l'espace des sociétés, sous la direction de Jacques Lévy et Michel Lussault, Événement.
[6] Nora Pierre, *« Le retour de l'événement »,* in Le Goff Jacques et Nora Pierre (dir.), *Faire l'histoire*, TI, Paris, Gallimard, 1974, p 452-
[7] Manuel Castells s'est attaché à décrire depuis une vingtaine années les liens qui existent entre l'extension de l'espace des flux et des réseaux et l'affirmation de lieux de polarités fondés sur la proximité et des identités résistances ou d'identités projets dans le cadre de la globalisation.

*L'évaluation ?*

Mais qu'est-ce qu'une évaluation ? Comment évaluer et qui évalue, la désignation de l'évaluateur[8] ? Pour le terme d'évaluation, celui-ci apparaît en 1366 - de *evaluacion* venant du verbe évaluer. Pour le verbe, évaluer, celui-ci apparaît en 1365 – de *eavaluer* ou d'une variable *avaluer* (1230). Il s'agit en fait d'une action dont on parle, celle d'évaluer. C'est en fait une appréciation, une détermination, une estimation par une expertise issue parfois de la comparaison ou d'une grille d'analyse. Généralement, on comprend l'évaluation par une volonté de l'évaluateur de faire un inventaire, une majoration (dans son sens premier) ou de donner une approximation. Il ne faut pas oublier aussi que le terme renvoie à la notion d'échelle dans le temps et l'espace. Enfin, l'évaluation dans son deuxième sens, c'est la valeur, la quantité évaluée, les synonymes du Grand Robert de la langue française font apparaître les termes de mesurer, prix, valeur.

N'oublions pas que ce sont les professionnels dans les domaines juridiques et fiscaux qui se sont accaparés une bonne partie de l'emploi du mot : évaluation budgétaire, détermination des dépenses et recettes à inscrire au budget ou bien l'évaluation d'office et estimation de bénéfices, de profits, effectués par l'Administration sans procédure contradictoire[9]. Mais de nos jours, l'évaluation est de plus en plus incertaine, car rappelons-nous encore la citation de Charles Pegy, *« il ne faut donc pas dire seulement que dans le monde moderne l'échelle des valeurs a été bouleversée. Il faut dire qu'elle a été anéantie, puisque l'appareil de mesure et d'échange et d'évaluation a envahi toute la valeur qu'il devait servir à mesurer, échanger, évaluer »* [10]. L'évaluation n'est pas une analyse de ce qui s'est passé, mais le fait de donner une valeur aux éléments faisant corps et sens pour l'événement après sa tenue.

Il faut donc identifier les changements, les évolutions et donner une échelle de correspondance. Soit une action d'évaluer pour donner une valeur à l'objet étudié. Une valeur positive, négative, neutre soit un aspect quantitatif ou qualitatif intrinsèque. Ne l'oublions pas, nous vivions dans une société de l'audit[11] et du *« grenouillage »* si l'on prend un dictionnaire du vieux français[12].

---

[8] D'ailleurs le terme d'évaluateur a été inventé par Proudhon au milieu du XIX ème siècle. Cf. son livre sur *« une Exposition permanente »* en 1867.
[9] Le Grand Robert de la Langue Française, TVI, 2001 p. 233
[10] Pegy Charles, *« La République »,* Paris, Seuil, 1900, p. 351
[11] C'est en fait, l'obsession du contrôle de Michael Power, *La société de l'audit*, Paris, E. et S., 312 p.
[12] J'emploie ce terme qui est à l'origine du terme magouille, magouiller… pour montrer que l'audit a ses limites malgré les dérives et corrections constatées ces dernières années. On s'aperçoit que les comptes des entreprises et leurs prévisions sont souvent fausses. Sans parler des comptes et prévisions du comité olympique grec pour recevoir les JO. La candidature et l'organisation des Jeux olympiques et Paralympiques d'été d'Athènes 2004 permettent de prendre en compte le rôle essentiel de l'audit ante événement dans le cadre du portrait économique du pays hôte des olympiades. En effet, l'événement olympique de 2004 est issu d'une longue négociation entre le CIO et l'Etat Hellène. Quand la capitale grecque propose une première candidature olympique en 1996 – attribuée à la toute fin à Atlanta (États-Unis) – le Comité International Olympique (CIO) soulève la question de la solidité économique et financière du pays [CIO, note de synthèse des évaluations des olympiades 1996, 1992]. De fait, pour remédier au portrait financier et économique du pays, les dirigeants grecs ont maquillé leurs comptes publics pour entrer dans l'euro, en 2001 et obtenir finalement l'accueil des JO en 2004. La dissimilation d'emprunt toxique sur les conseils de la banque américaine Goldman Sachs tourne au fiasco pour l'un des plus pauvres États de l'Union européenne, avec la complicité de bureau et d'agence d'audit internationale pour les emprunts sur les marchés extérieurs comme par exemple JP Morgan donnant quitus à l'euro et à l'olympiade comme source de croissance. L'ensemble de ces évaluations trompeuses et mécanismes financiers hasardeux entraînèrent une illusion auprès de la population et un endettement record aggravant la situation du pays avant son quasi-défaut. On n'oubliera pas non plus la très grande corruption du système politique grec.

Il faut être réaliste. Par exemple : un maire d'une grande ville qui veut organiser un événement touristique pour la première fois ou renouveler un événement qui connut un certain succès des décennies auparavant. Celui-ci veut légitimer son action et par là même instrumentaliser ses ordres. Il décide de faire appel à une société d'audit. Il lui commande un rapport : le rapport peut-il être véritablement objectif et présenter une vision altérée de ce que veut le maire. Le maire va-t-il s'en servir - le publier ou non - suivant les résultats…

À l'occasion du choix de la ville hôte des Jeux olympiques de 2012, aucun des rapports consultés ne furent négatifs, tous furent dans la même lignée[13]. Il existe des cabinets d'audit honnêtes et d'autre plus soucieux d'aller dans le sens de la mode ou de leur client. Le cabinet honnête alertera le demandeur, l'autre rentrera dans un jeu plus complexe de présentation positive des olympiades tout en laissant de côté ou en annexe de dossiers les nuances d'un méga-événement comportant toujours une part d'incertitude. Les conséquences peuvent être importantes et sauver les apparences sur le moment comme pour le mandat présidentiel de Jacques Chirac comptant sur la candidature olympique de Paris pour redorer son image et son mandat est un exemple parmi tant d'autres dans la longue histoire de ces manifestations géantes. [Impression française de l'auteur de l'article concernant la campagne pour Paris 2012]. D'ailleurs, concernant les calculs d'évaluation ante événement, il est très important de spécifier la méthode de calculs ou d'une recherche d'ordre quantitative ou qualitative.

En fait, on analyse et recherche une sorte de « paysage économico-socio-culturel » produit par l'événement et son financement adéquat par une participation de l'ensemble des forces vives de la nation accueillante. Mais bien souvent, les retombées qualitatives sont difficiles à prouver ou à classifier. Néanmoins, il est possible de proposer une évaluation de ce bien-être, une aménité supplémentaire qui fait la différence. L'évaluation post-événement est beaucoup plus légitime. Elle comprend l'observation de l'événement et ses suites. Le tout est de cerner l'évolution entre l'avant et l'après concernant les dimensions touchées par le projet. Par contre, le recours à des évaluations avant la tenue même de la manifestation rentre dans la politique de justification de l'organisation de celui-ci, une arme de propagande et de communication marketing que l'on peut très bien comprendre à retrouver dans un dossier de candidature sans toutefois avancer qu'il s'agit d'une vérité absolue [une estimation].

*Le prétexte*

Dans le sens premier de la définition du terme prétexte, on aborde le fait qu'il s'agit d'une raison alléguée pour dissimuler le véritable motif d'une action. Puis dans son deuxième sens, le prétexte est ce qui permet de faire quelque chose, une occasion. En fait pour évaluer un événement touristique de taille tel que les Jeux olympiques ou une exposition internationale - universelle, il faut prendre en compte le fait qu'il s'agit *d'une invocation*, d'un *motif mis en avant* souvent fabriqué, une raison alléguée qui cache la véritable cause d'une action, d'une attitude, d'une volonté urbanistique et sociétale pour Barcelone, notre terrain d'étude. Les Jeux ou les expositions universelles servent très souvent de justification - prétexte.

---

[13] Bien que l'inflation des villes candidates tend à montrer l'attractivité de ces événements, le poids des villes asiatiques dans les prochaines années va devenir prépondérant et l'endettement de l'Occident une limite à l'organisation des olympiades en Europe et Amérique.

Généralement, le prétexte est ce qui sert de point de départ pour une cause non révélée ou connue de tous comme le fait de refonder une ville ou de requalifier un quartier ou une région urbaine dans le cadre *de très grands projets urbains issus de l'organisation de très grands événements touristiques.* Il faut aussi souligner dans les argumentaires des décideurs politiques, sportifs et culturels, la sempiternelle quête de l'infrastructure lourde faisant le lien entre la société et la ville *« je vous apporte cette infrastructure symbolique »*.

L'alliance des trois entraîne une réaction en chaîne parfois bien difficile à maîtriser. Une manifestation est invoquée, elle est *un prétexte*, une évaluation liée au principe d'espérance d'un projet est mise en avant (Bloch, 1947), c'est l'évaluation ante événement. L'important réside dans ce qui restera de l'événement souvent dans le domaine de l'infrastructure ou des pratiques culturelles ou sportives. Souvent à cette occasion, c'est l'utopie qui est en marche avec le degré zéro des contraintes à la fois budgétaires, mais avec parfois des groupes de pression ou des adversaires invoquant de plus en plus les aspects écologiques. Des cartes mentales se matérialisent par des cartes à projet souvent dans la presse quotidienne qui deviennent à leur tour des mises en scène et en récit de la ville pour le meilleur et parfois le pire en cas d'échec ou de déficits colossaux. L'infographie est aussi au rendez-vous avec le jeu de maquette numérique et d'infographie pour justifier une communication et un marketing du projet(figures n°1).

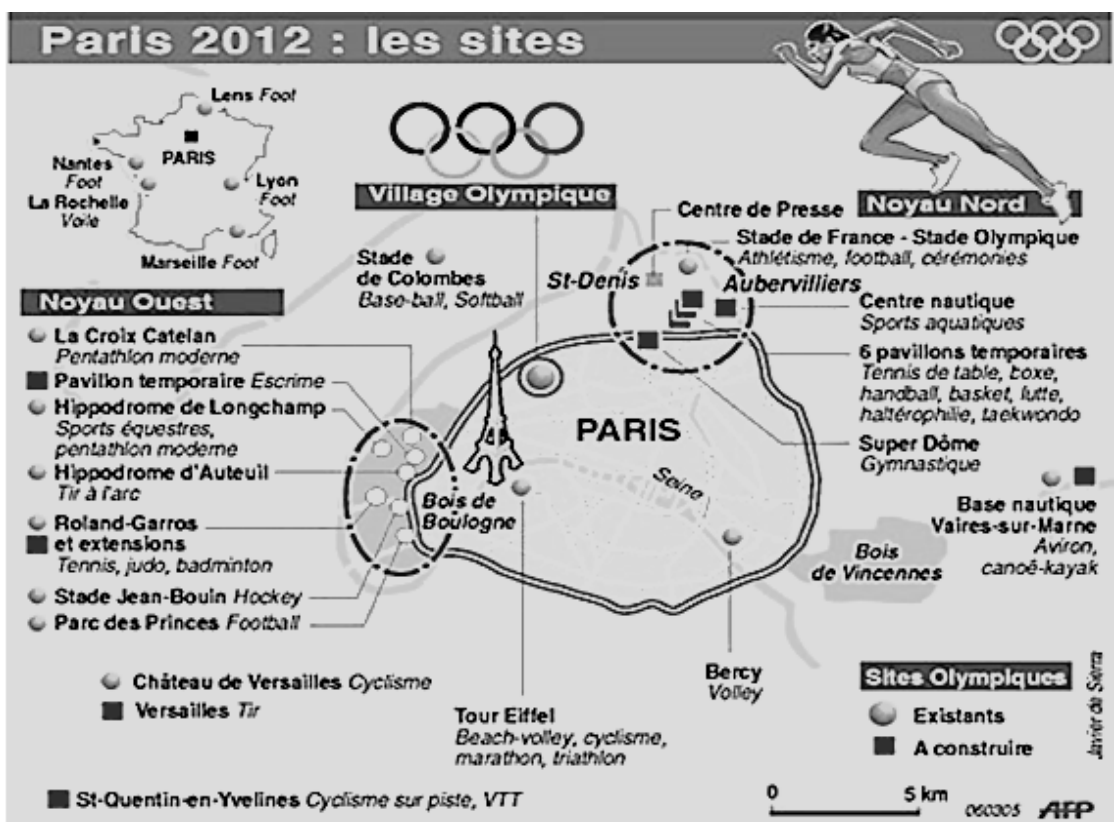

Figure 1 : Paris 2012, les sites. Comité d'organisation olympique pour AFP, 2007. [ La carte comme propagande].

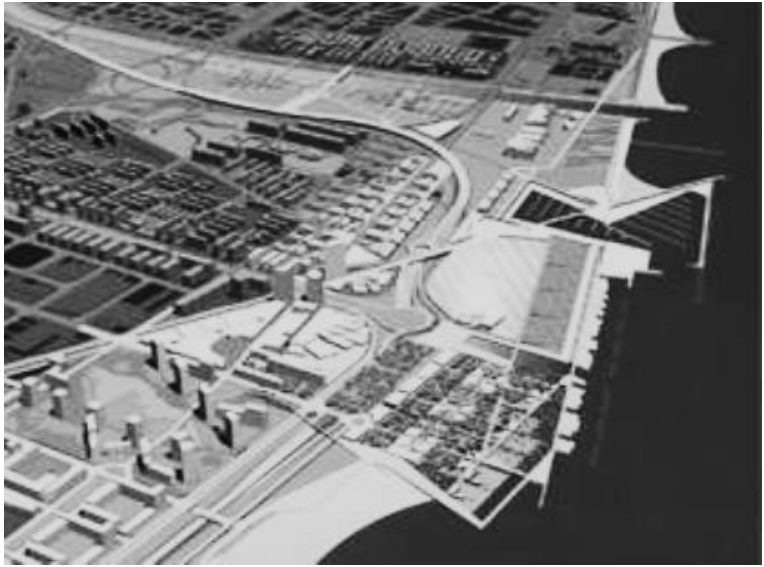

Figure 2 : Plan directeur Forum universel des cultures 2004 - Llevant photographie aérienne, premier projet, Infrae2004, 2000/2002 magazine municipal et presse régionale.

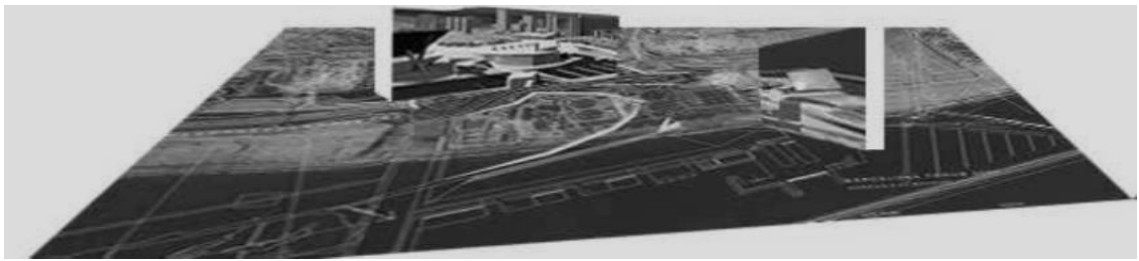

:
Figure 3 : Première présentation sur le site BCN du projet Forum, schéma interactif montrant les principaux espaces symboliques. www.bcn/estrategiasurbanas/forum-origins.cat, au 11/12/2002 [L'infrastructure et le digital comme marketing et communication urbaine.

## 2. Les différentes approches de l'évaluation touristique : méthodologies et proposition de grille générale d'analyse temporelle, spatiale et sociétale

Une question se pose de contextualisation, périodisation, méthodologie et de comparaison d'un événement. Différents bureaux d'études dans le monde essayent de modéliser les effets de Jeux olympiques ou d'exposition universelle et internationale, mais parfois les doutes et incertitudes subsistent.

*De la régénération urbaine*

L'après Jeux olympiques ou expositions universelles ou internationales devient une réalité difficile à cerner, bien que le nombre de touristes soit possible à comptabiliser, la dotation en infrastructure aussi par une simple comptabilité qui peut devenir complexe en cas de service touristique non performant dans la récolte de statistique et surtout du rapport, coût/qualité/prix/durabilité de l'équipement public ou semi-privé hérité dans la trame urbaine. Le capital humain et la formation de spécialistes socio-culturels ou sportifs post événement peuvent s'entrevoir, de même que la volonté de régénération urbaine associée à l'utopie d'intervention.

Quant à la réputation internationale, le bruit médiatique la couverture positive de l'événement par tous les organes de presse ; cette dernière n'est pas aisée à prendre en compte. Quant aux risques, ils sont très importants : la maintenance et la rentabilité des infrastructures construites sont un enjeu primordial qui peut conduire à la perte d'image positive et de déficits conséquents ainsi que la fuite de concitoyens préférant consommer ailleurs ou proposant de nouvelles habitudes après des années de travaux intenses en ville[14].

La régénération urbaine revient encore sur le devant de la scène dans son côté spectaculaire et visible. L'étude des hypothèques de Vancouver pour les Jeux d'hiver de 2010 est la plus aboutie : elle rappelle l'importance de la vie des infrastructures et des espaces publics après l'événement. Il en va de même pour les expositions internationales et universelles avec essentiellement une focalisation sur l'enceinte d'exposition. Sydney est un exemple paradoxal d'échec et de réussite à la fois. Sur le court terme, les Jeux olympiques sont considérés comme une réussite financière et médiatique. Sur le moyen et long terme comment les aménageurs ont-ils pu croire que les infrastructures sportives projetées en périphérie lointaine de la ville allaient connaître un taux de remplissage quasi identique à celui de la période des jeux ? Cette opération devait amener des promoteurs immobiliers et des entreprises. Le taux d'utilisation de ces infrastructures sportives reste en dessous des prévisions proposées qui ont été vantées. Le bilan des billets et droits de télévision est une chose, par contre le bilan sur le long terme peut devenir politiquement incorrect, car il engage les finances publiques.

*La stratégie d'accompagnement*

Il est un fait, la stratégie d'accompagnement des espaces publics créés devient l'un des principaux objectifs des acteurs de la manifestation. Si malheureusement, elle n'est pas prévue, on aboutit systématiquement au catastrophique endettement public de villes pour les Jeux de Montréal de 1976 et surtout à une plus large mesure Séville 1992 exposition universelle. Ils représentent des contre-exemples au même titre que les JO d'Athènes 2004. Que doit-on prendre en compte comme canevas d'évaluation ? Les tableaux indicatifs donnent des pistes et des avancées dans les domaines qu'il faut regrouper prenant comme exemple les travaux de la Fondation Jacques Cartier à Lyon. Deux facettes se déploient dans ces très grands projets de régénération urbaine, ce qui est comptabilisable et ce qui est de l'ordre du symbolique : *« l'impact économique des grands projets doit être mesuré d'une façon différente selon que l'on raisonne à la manière des acteurs privés ou des acteurs publics. Pour les uns, l'investissement doit être rentable en moins de quinze ans ; pour les autres, les investissements à très long terme ont un sens particulier, qui repose sur des critères non exclusivement économiques, mais aussi symboliques ou relatifs à la cohésion sociale »*[15].

---

[14] Société canadienne d'hypothèques et de logements, Imprimé au Canada. Réalisation 2006 SCHL 02-11-06 en ligne. Cf. bibliographie et principaux audits de bureaux d'études sur les impacts des méga-événements. La liste est pléthore : le sérieux canadien est à souligner peut-être du fait que, comme pour les JO américains, la demande de bénéfices à court terme est obligatoire comme pour la destruction du stade d'Atlanta remplacé par un centre commercial, ou de Los Angeles 1984 avec la destruction de bâtiments éphémères dans leur quasi-totalité (la ville ne comptant que sur les droits TV et vente de billets, merchandising). Le comité olympique américain exigeant un financement quasi privé (hormis la sécurité, police…).

[15] Shearmur Richard, Polese Mario, « Pourquoi Toronto a surclassé Montréal au sommet de la hiérarchie urbaine du Canada ? L'impact des différences culturelles sur la dynamique spatiale des services supérieurs. », in *Géographie, espace et société*, 2003, n°5, 3-4, pp. 399-420.

Pierre Bouchard, ajoute qu'en outre de la dichotomie, c'est la création de nouvelles pratiques de gouvernance dont il est question, principalement dans des configurations libérales comme au Canada ou en Espagne. : *« les limites budgétaires de la ville ne permettent pas de transgresser les normes, les règles et les standards. Toutes ces pressions exigent de la part des services publics de se dépasser, de revoir leurs façons de faire et de s'interroger sur la nécessité de maintenir certaines normes et contrôles. Cette remise en question ouvre la porte à l'innovation et au progrès, un peu comme si le grand projet était un vaste laboratoire permettant de revoir (nos) méthodes de gestion. »*[16].

De surcroît, il faut *« gérer l'incertitude »* comme le rappelle Jean Frébault : *« Il faut faire face à des aléas qui peuvent être soit positifs et sur lesquels il faut savoir rebondir, soit au contraire négatifs – changements politiques, conflits institutionnels, conjoncture immobilière défavorable – et qui peuvent casser la dynamique du projet »* (2002). La flexibilité et la réversibilité sont requises et garantissent un succès. Néanmoins la tâche est ardue, car s'insère un autre facteur important : la cohésion ou la mixité des acteurs (publics, privés, associations, mécènes, sociétés d'économie mixte) permettant une conduite optimum du projet jusqu'à son achèvement. Point commun à tous ces projets : c'est l'acteur public qui dirige le curseur : *« Effet d'impulsion initial de la part des acteurs publics »* et pas uniquement dans le choix du site emblématique à requalifier ou dans le fait de sortir d'une situation de crise exceptionnelle. Avec Toronto 2010 et ses 40 kilomètres de *waterfront* urbain et de côte lacustre, nous avons une société privée dont le capital est détenu par la municipalité, le gouvernement et l'État fédéral. Avec la mondialisation, les grands projets ne sont pas systématiquement pourvoyeurs de richesse, de développement et d'emploi stable. On peut tout à fait prendre en compte les impacts négatifs : *« sur l'emploi, les déséquilibres sociaux ou la dégradation de l'environnement, du non-investissement de métropoles dans la sphère sociale confrontées à la sévère compétition internationale »*. Il reste alors la qualité architecturale et la conception d'espaces publics présentant une valeur ajoutée non négligeable. Au plan foncier et social, c'est toujours la gentrification qui est au rendez-vous ou les faillites retentissantes de groupes immobiliers non préparés aux aléas[17].

### *De la recherche des temps d'analyse pour une évaluation*

La dimension + 20 ans est essentielle pour bien juger de l'impact de ses très grands projets urbains, notamment pour l'impact sur les transports et une meilleure mobilité comme lien social et moyen d'augmenter l'attraction touristique[18]. Elle prend en compte la pérennité des bâtiments et la création d'une identité et d'un nouvel héritage urbain, sans contestation possible de par le pouvoir palimpseste de ces manifestations qu'elles engendrent ainsi que de son évocation. Le résultat recherché est la création de représentation forte sur des fondements économiques et sociaux solides. Il peut aussi se voir une alliance informelle entre le secteur des forces vives de la nation et des organisateurs pour permettre une mobilisation générale de la société urbaine.

---

[16] Responsable du bureau des grands projets à Montréal.
[17] Colloque *« Les grands projets de revitalisation urbaine et Métropolitaine »*, les XIV èmes Entretiens du centre Jacques cartier. 3 et 5 décembre 2001. Ensemble des propos recueillis par Pierre Gras de la revue Urbanisme.
[18] Philippe Bovy Professeur EPFL, Directeur du Laboratoire de Mobilité et Développement territorial de l'EPFL et Expert transport auprès du CIO a fait de cette question un axe principal de ces recherches. *Jeux Olympiques : quelles évolutions dans la gestion d'un méga-événement, notamment celle des transports et des mobilités ?*

On instrumentalise très souvent des exemples de villes comme Barcelone 1992 JO et Séville 1992 pour les Expositions, pour arriver à présenter au grand public un projet attractif et réussi. Le cas de Barcelone est intéressant : il revient dans tous les dossiers de candidatures des JO y compris pour la dernière élection de Rio de Janeiro avec l'ensemble des participants lors des séances de présentation des projets urbains mettant en adéquation port de Barcelone / Port de Rio de Janeiro, un nouveau waterfront.

Ce procédé est aussi une demande venant même du CIO pour présenter un projet non pas coûteux, mais ambitieux pour une ville, une nation, une région. Barcelone revient très souvent et l'étude de l'UAB de Père Duran comme modèle de propos et d'analyses des JO de Barcelone 1992. Néanmoins, la capitale catalane est partie de très loin, car on constate une quasi-absence de périphérique, de quartier touristique de front de mer, d'infrastructure de câblage, de place publique, d'immobilier et logements pour tous. Les autorités catalanes proposent un projet urbain novateur, la capitale catalane dispose d'une interface maritime, d'un port et de nouvelles plages pour un patrimoine culturel qui ne demande qu'à être réactivés. Souvent, le bond en avant de Barcelone est trompeur, car il s'inscrit sur plus de 20 ans de réflexion urbaine avec un événement comme levier urbain et de propagande mondiale d'une image renouvelée. Par contre des contre-exemples sont possibles comme lors de la dernière campagne électorale à Paris avec Christine de Panafieu proposant une Exposition universelle pour Paris. Les sondages du parisien et les réactions des habitants ainsi que des alliés politiques de la candidate furent en partie négatifs, car il est vrai qu'une exposition universelle n'est pas une olympiade, ni dans sa tenue ni renommée et financement par le sponsoring [et pourtant au plan temporel, elle se déroule pendant 6 mois où les médias peuvent revenir plusieurs fois sur la métropole et les différentes attractions et pavillons présents].

L'évaluation peut aussi être un moyen de légitimation, une stratégie de communication pour concrétiser un projet. C'est le cas pour les Olympiades et les Expositions internationales de Barcelone, leur souvenir appelant à d'autres répliques. La nature d'une évaluation peut en soi proposer une stratégie de conceptualisation de l'événement et de ce qu'il peut rapporter sur le plan du marketing urbain : sortir un chiffre - et tout devient possible - avec en plus le ciblage d'un quartier en difficulté comme lieu du village olympique par exemple. On intègre la capacité à proposer des évaluations projets, bilans ou modèles. Il faut alors connaître le moment où l'on pourra en tirer des réponses satisfaisantes et non variables concernant la détermination d'une évolution (tableau n°1) et les cinq temps de l'évaluation d'un projet de méga-événement (tableau n°2).

**Tableau n°1 : À la recherche des temps d'approche pour une évaluation du méga-événement.**

| TROIS TEMPS D'APPROCHE POUR l'ESTIMATION | QUATRE AXES MAJEURS D'ESTIMATION (Jeux olympiques /Exposition internationale - universelle) (A partir de quel moment peut-on analyser les effets de ces événements ?) | | | |
|---|---|---|---|---|
| | **ÉCONOMIQUE** | **TOURISTIQUE** | **SPORTIVE/CULTURELLE** | **URBANISTIQUE** |
| **Court terme + 1 an** | **Déficit budgétaire ou non de l'organisateur** | **Nombre de touristes supplémentaires et leur origine** | **Pratique sportive culturelle Fréquentation des musées, comportement sain et vente d'article de sport ou biens culturels** | **La fréquentation des nouveaux espaces ?** |
| **Moyen terme + 7 ans** | Quels impacts sur la valeur ajoutée, sur les emplois créés, entreprises et nouveaux marchés ouverts. | Évolution du nombre de nuits/hôtel sur une dizaine d'années. Une image à vocation mondiale ? Loisirs / affaires | Phénomène d'amélioration de la santé publique par la pratique d'un sport régulier sur le moyen terme. (Statistique sanitaire) Culture retrouvée ou fréquentation de bibliothèque. En lien avec la lutte contre la pollution urbaine ou politique éducative | Un environnement renouvelé, dépollué ? |
| **Long terme + 20 ans** | Un effet suramplificateur au plan économique par la réunion des trois autres variables, touristique, sportive/culturelle/ urbanistique. | L'image de la ville olympique ou d'exposition : dynamique et culturelle reconnue Une stabilité ou croissance dans le nombre annuel de visiteurs. Loisirs/ Affaires | Pratique devenant réflexe culturel. | Un héritage urbain ? Un événement identitaire ? Des aménités d'importance permettant la relocalisation d'entreprises. |

**Par Patrice Ballester**

**Tableau n°2 : Les cinq temps de l'évaluation comme projet - bilan-modèle**

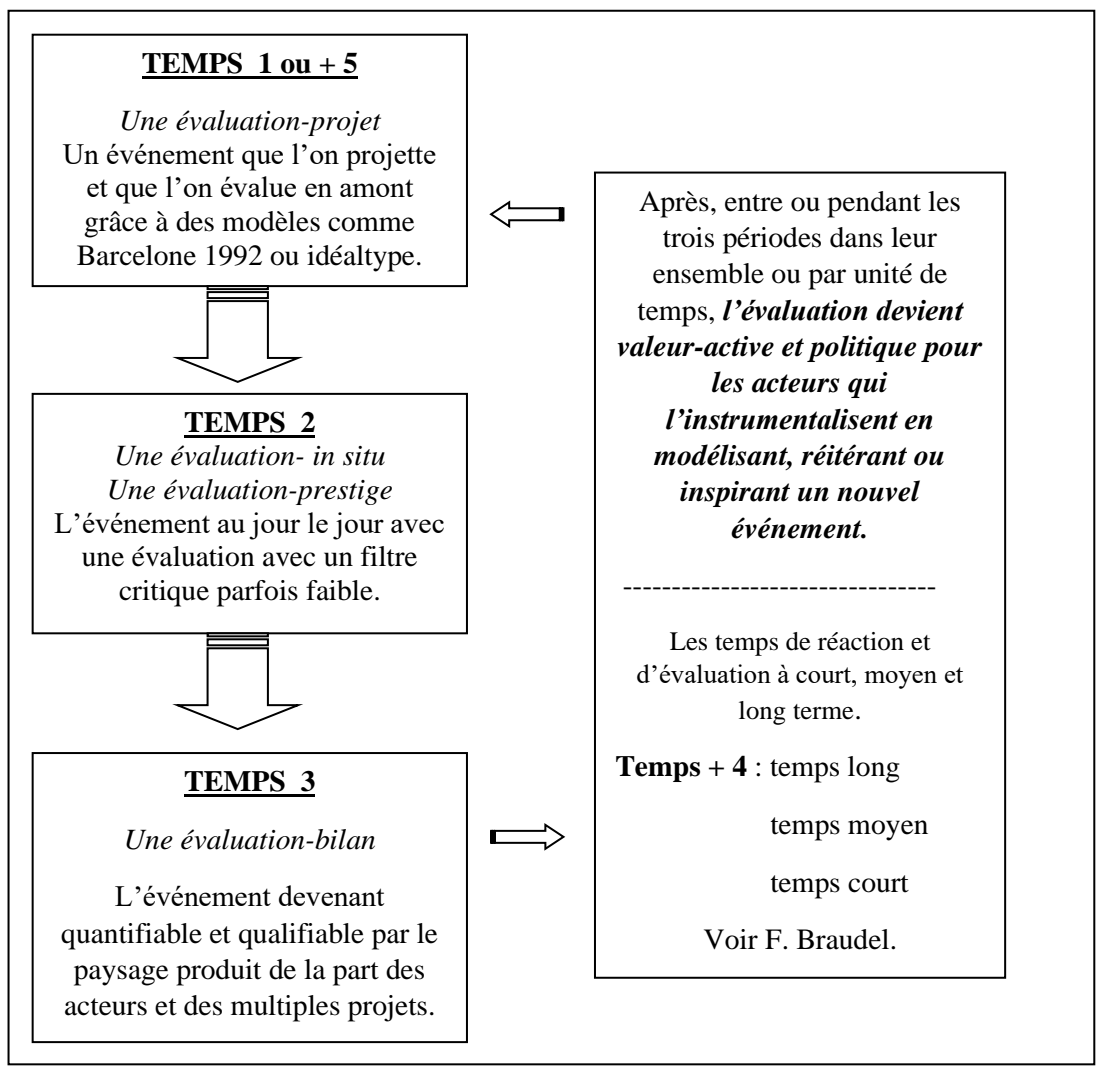

**Par Patrice Ballester**

Nous estimons qu'il n'est pas opportun de revenir sur les calculs d'évaluation ante événement, ils ne sont pas probants et ne donnent que quelques indications sur la valeur accordée à l'événement par les acteurs avec les grands objectifs à atteindre. Le constat produit par l'événement et son financement adéquat par une participation de l'ensemble des forces vives de la nation accueillante est un objectif louable. Hélas, l'image, les retombées qualitatives sont souvent difficiles à prouver ou à classifier.

Néanmoins, il est possible de proposer une évaluation de ce bien-être avec les aménités nouvelles géographiques, culturelles, identitaires et symboliques (éco-techno-symbolique) qui font la différence dans le temps moyen ou long. On peut estimer que les dimensions sont aux nombres de 11 dont une qui regroupe l'ensemble dans une sorte de visions de « *l'économie du bien-être* » (Arrow, 1957). C'est un regard mondial, une vision globale citoyenne, des acteurs, des observateurs étrangers et des experts scientifiques et financiers estimant que telle métropole est digne d'intérêt pour des investissements futurs. Le tout est de savoir ce que l'on a voulu faire et ce qui a été réalisé dans la limite des moyens financiers, avec la prise en compte des effets directs et indirects. On retrouve ce que la société peut supporter ou non dans une ambiance festive pour les Jeux olympiques et les expositions internationales universelles, dans le cadre d'un projet à décloisonner et à réinterpréter.

Table 3.1 – Key economic benefits and costs of the Games

| | Benefits | Costs |
|---|---|---|
| Pre-Games Phase | Tourism<br>Construction activity | Investment expenditure<br>Preparatory operational costs (including bid costs)<br>Lost benefits from displaced projects |
| Games phase | Tourism<br>Stadium & infrastructure<br>Olympic jobs<br>Revenues from Games (tickets, TV rights, sponsorship, etc.) | Operational expenditure associated with Games<br>Congestion<br>Lost benefits from displaced projects |
| Post-Games phase | Tourism<br>Stadiums & infrastructure<br>Human capital<br>Urban regeneration<br>International Reputation | Maintenance of stadiums and infrastructure<br>Lost benefits from displaced projects |

Figure 4: The economic impact of the Olympic games Price Water House Coopers European Economic Outlook June 2004, p.18)

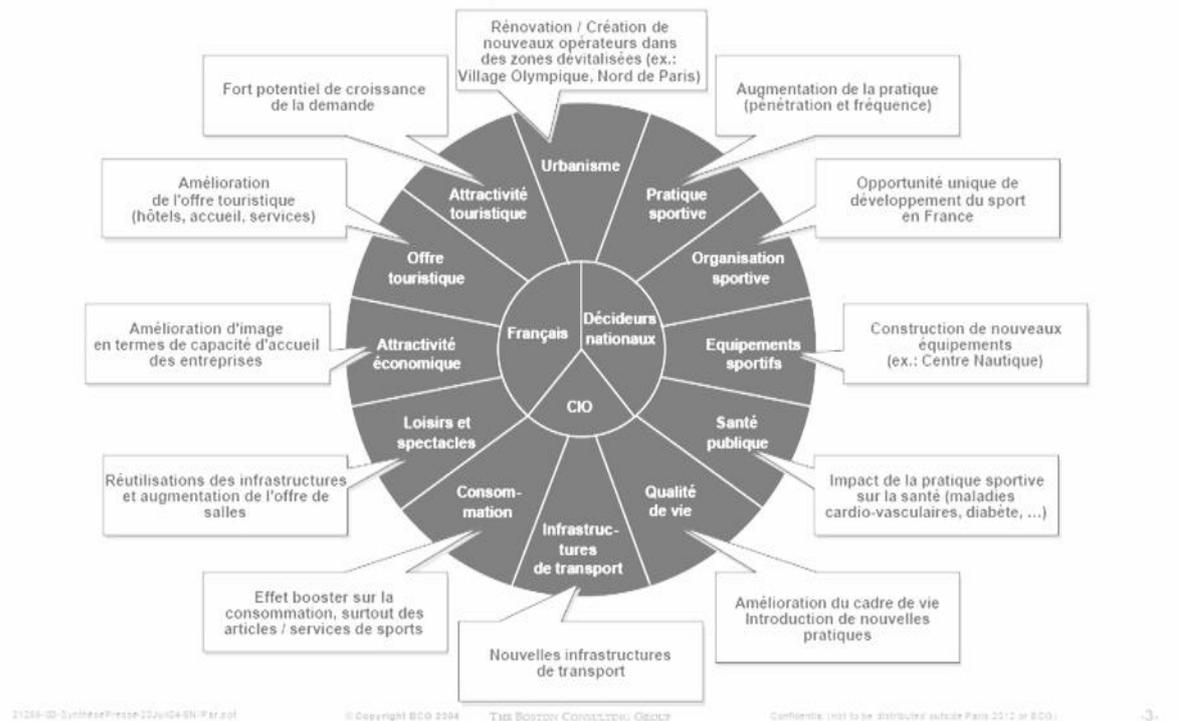

Figure 5 : Boston Consulting Group, identification des critères d'évaluation et d'impact des JO de 2012 Paris. p.11.

À savoir, par Patrice Ballester :

**1-**Comment les décideurs appréhendent-ils la ville-région-pays dans une réflexion d'ensemble concernant l'organisation de l'événement. Le choix d'une zone à revitaliser, l'importance de la régénération urbaine, d'un axe urbain, d'un quartier, d'une friche, par la création de nouveaux opérateurs urbains ou procédés urbanistiques récents « exemple : le droit à l'urbanité barcelonaise », l'inflation des programmes sociaux immobiliers. Une société se met en ordre pour être prête au jour J du commencement de la manifestation. Souvent, les services d'urbanisme et de construction publique se restructurent comme à Barcelone et se projettent au-delà de la manifestation même pour enclencher l'effet d'accompagnement comme pour la création d'un Office de Tourisme moderne pour la capitale catalane.

**2-**Le transport des visiteurs et la cohabitation avec les locaux, élément essentiel déjà pris en compte lors des Expositions parisiennes ou barcelonaises depuis très longtemps. Création de nouvelles infrastructures de transports pour permettre une fluidité des déplacements et un accueil en nombre des touristes. Ici l'héritage urbain est probant avec quelques déconvenues comme certaines gares ferroviaires routières du sud de l'Italie pour le mondial de football de 1992.

**3-**La construction de nouveaux équipements sportifs, culturels, médiatiques et artistiques. La réutilisation des infrastructures et l'augmentation de l'offre des salles de spectacles, musiques, compétitions sportives ou réunions, congrès, colloques. Le taux de remplissage, la réception de la qualité de l'infrastructure et le coût de l'entretien sont primordiaux.

**4-**Impact de la pratique sportive sur la santé (maladies cardio-vasculaires, diabète…), mais aussi impacts sur l'éducation et la pédagogie du savoir lors d'une exposition internationale, universelle (prise en compte de l'environnement, danger de la pollution par exemple dans un cadre pédagogique).

**5-**Amélioration du cadre de vie. Introduction de nouvelles pratiques urbaines en relation avec les nouveaux espaces issus de la requalification urbaine. De nouveaux paysages perçus et vécus autrement. Actuellement, le principe structurant du développement durable.

**6-**Spectacles, musique et fêtes dans les rues de la ville = unité de temps, de lieu et d'espace avec la manifestation. Architecture de l'éphémère et durabilité du bien-être social. Des olympiades culturelles à la thématique culturelle d'une Exposition universelle ou internationale. Un exemple, *Freedom for Catalonia*, permet de prendre en compte l'importance de la capacité à adosser un projet de reconnaissance culturelle à un projet olympique, sportif et projet urbain renouvelé.

**7-**Amélioration de l'image : (1) en termes de gestion opératoire d'un grand projet événementiel, (2) en termes de capacité d'accueil des entreprises, (3) en termes de promotion de nouvelles ou anciennes valeurs sportives ou culturelles.

**8-**Amélioration de l'offre touristique : (1) capacité hôtelière, (2) capacité à offrir de nouveaux services, (3) amélioration de l'accueil (exemple, langues parlées : anglais, chinois, japonais, russe, espagnol…) grâce à l'offre démultipliée et (4) alliance tourisme de loisirs et d'affaires renouvelés.

**9-**Image positive de la ville, son rayonnement national, international globalisant, entraînant un effet de mode sur la demande. Mesurer le bruit médiatique de l'événement par les nouveaux réseaux sociaux numériques.

**10-**Développement : (1) du sport ou de la culture par la création d'entreprises innovantes ou « surfant » sur l'événement, (2) un effet *« suramplificateur »* sur la consommation d'articles sportifs/services de sport, mais aussi sur la consommation de biens culturels. Pour les JO, augmentation de la pratique sportive (taux de pénétration et fréquence, plus nouvelle pratique sportive). Pour une exposition internationale, pénétration des idées maîtresses véhiculées par les pavillons de l'exposition, pratique des musées et achat/mode de consommation en lien. (3)

**11-**Effet supposé ou non *« France 98 »*, à relativiser depuis, fête identitaire, célébration nationale, baromètre psyché. Néanmoins un nouveau paysage émerge de cet événement : globalement, le perçu et le vécu deviennent vision et réalité alternative parfois. Une péréquation de l'ensemble se produit.

# L'ÉVALUATION DE L'ÉVÉNEMENTIEL - LES 11 DIMENSIONS DE L'ÉVALUATION (JO – EXPO)

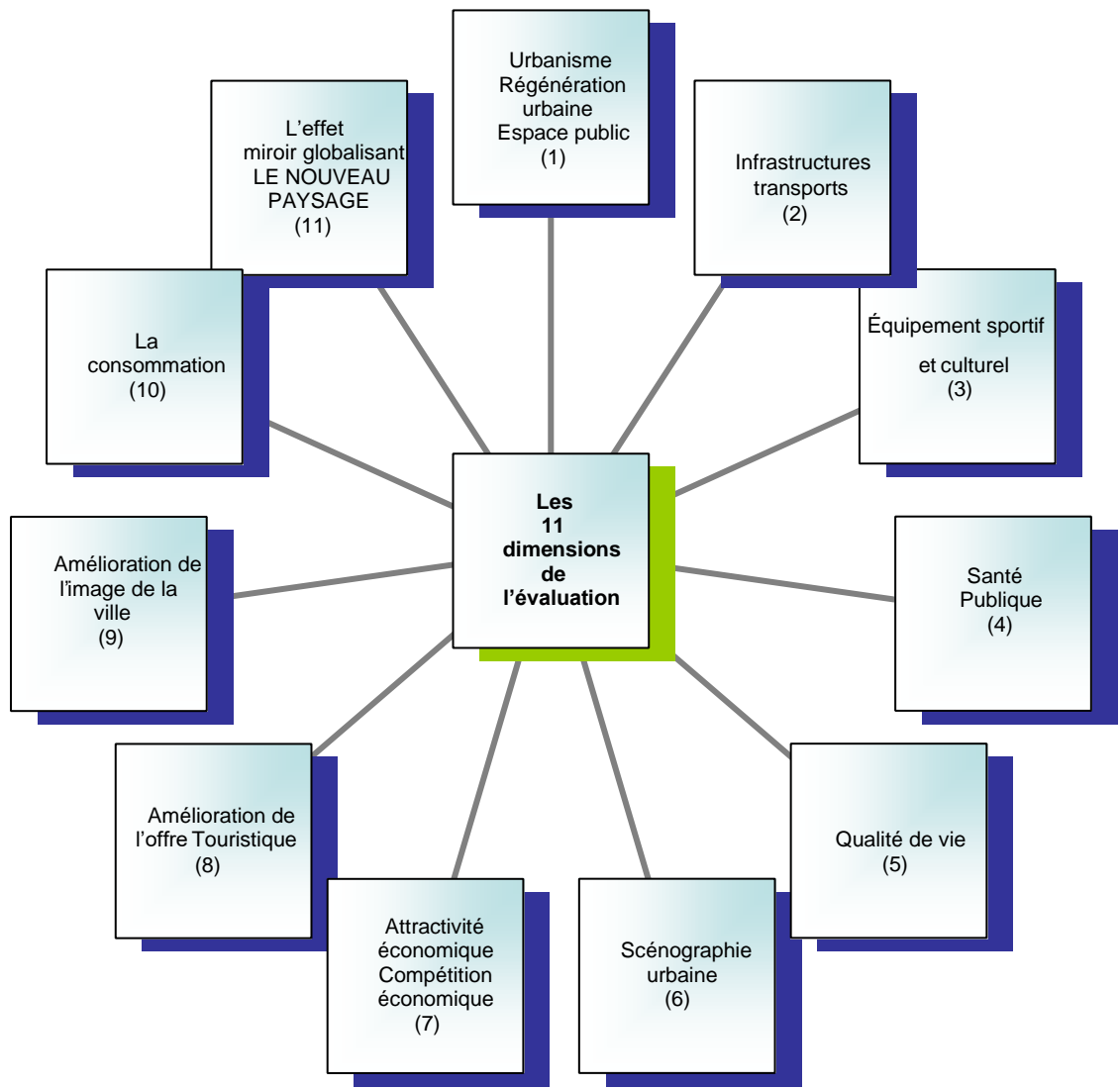

Par Patrice Ballester

## 3. Étude de cas sur Barcelone : mondialisation, événementiel et évaluation

Les expositions internationales et universelles barcelonaises émanent d'une civilisation occidentale dont la pratique vise à exposer son savoir-faire. Elles sont entrées dans le patrimoine commun européen culturel et s'exportent à l'étranger et dans les pays émergents depuis leur création.

De plus, elles ont l'avantage de nous permettre d'analyser, pour chaque époque, leur impact sur la trame urbaine barcelonaise ; impact en rien insignifiant, car au contraire significatif dans l'évolution de leur rôle et fonction. Une matrice, certes, mais aussi un révélateur d'appropriation de l'espace par l'Homme et d'imposition d'une idée civilisatrice et dominatrice d'une métropole qui cherche une croissance économique ou se rebelle sur le plan politique. La question du prétexte ou de l'occasion mobilisatrice des forces vives de la nation renvoie indéniablement à la part de l'utopie : on vend une redéfinition de la ville sur un concept mondialiste[18].

Dans les faits, les expositions internationales et universelles sont fréquemment caractérisées comme « *prétexte* » ou « *une débauche de moyens* ». Ceci conduit à une politique urbanistique involontaire, hasardeuse dans les premiers temps. Son imposition réfléchie, de forme urbaine, volontariste et dominatrice d'un territoire doit correspondre à l'image de ses ambitions premières ou de ses attentes en dotation d'équipements lourds, mais nécessaires pour une ville et son rayonnement culturel, sportif et touristique. Dans son article pour la onzième conférence de *l'International Planning History Society* à Barcelone, Javier Monclus, également architecte travaillant notamment sur des projets en vue de l'Exposition internationale de Saragosse de 2008, parle bien d'un phénomène *catalyseur* ou *régénérateur* pour les villes hôtes[19]. Néanmoins, on peut avancer quelques données et procéder à une requalification de ces événements, tout en proposant une typologie de travail et d'analyse portant non plus uniquement sur les impacts, mais sur la capacité qu'ont ces méga-événements à interroger toute une société. Nous comprenons alors que ces Expositions comme olympiades façonnent consciemment ou inconsciemment la ville et sa trame urbaine à divers degrés par un processus singulier impliquant : communication + production + distribution. L'évaluation propose de nombreuses dimensions : la question de l'identification des changements sociaux et économiques est primordiale. Elle comprend l'observation de l'événement et ses suites. Le détournement historique de l'évaluation de ces événements comme à Barcelone entre 1888 et 2004 tend à nous montrer modestes sur ce que l'on retient sur le très long terme d'un événement. La part du symbolique et du quantifiable deviennent essentielle. Pour le Forum 2004, comment prendre en compte le fait que l'on choisit le quartier le plus dégradé de la cité pour réaliser la manifestation ? C'est une part importante de l'évaluation dont les sommes engagées ne seront jamais *« remboursées »* par les entrées des 3 millions de visiteurs. Des manifestations qui endettent une ville deviennent emblématiques cinquante ans plus tard (figure n°5).

---

[18] Villette Agnès, « A l'Est de Londres. Avec les JO de 2012, l'Angleterre recouvre d'une nouvelle peau une de ses banlieues lépreuses. La fin de l'East End » in *Citizen K international*, 2007, pp.168-179 La plus grande friche urbaine d'Europe de plus de 400 hectares attend avec impatience les JO. À travers photographies et portraits d'habitants, le reportage montre une utopie-projet en marche, enclenchant pour les habitants des représentations multiples.

[19] Monclus Francisco Javier, « International Exhibitions and Planning. Hosting large-scale events as catalysts of urban regeneration », *IPHS Conference*, n°11, July 2004, p. 1-19. Qui a ce titre reprend l'expression de Mullin, Servant Takeda et Chaline concernant les termes de catalyseur et régénérateur. MULLIN, J.R. (1972)., *World´s Fairs and Their Impact Upon Urban Planning*, Monticello, Ill., Council of Planning Librarians, Exchange Bibliography, 303 SERVANT, C., TAKEDA, I.( 1996) *Study on the impact of International Expositions*, B.I.E., Paris.

Quand, les Sévillans entendirent parler d'une nouvelle exposition universelle en 1992, ils se rappelèrent des dettes de 1929, d'où la nécessité pour Madrid de garantir un minimum d'investissement. Pour Montréal, l'erreur de l'exposition universelle vient du manque d'entretien des bâtiments, car l'espace public hérité est fort attractif et symbolique pour la communauté québécoise, par contre, un autre type d'évaluation de cet événementiel rentre dans la capacité à proposer de nouveaux modes de constructions et d'aménagement urbain comme la sphère de Fuller. Événementiel touristique rime souvent avec nouveau mode d'habitation donc d'architecture et de technique nouvelle surtout pour les expositions universelles et internationales et à une moindre mesure pour les olympiades. Ils rentrent dans l'évaluation comme la pratique de nouveaux espaces publics. Enfin, l'équation-addition : 1888 (expo) + 1929 (expo) + 1992 (JO) + 2004 (expo) pour Barcelone se retrouvent souvent dans une mythologie urbaine auprès de la population forgée par la presse et les souvenirs des manifestations. C'est un point que nous allons approfondir dans le domaine de la manipulation de l'événement.

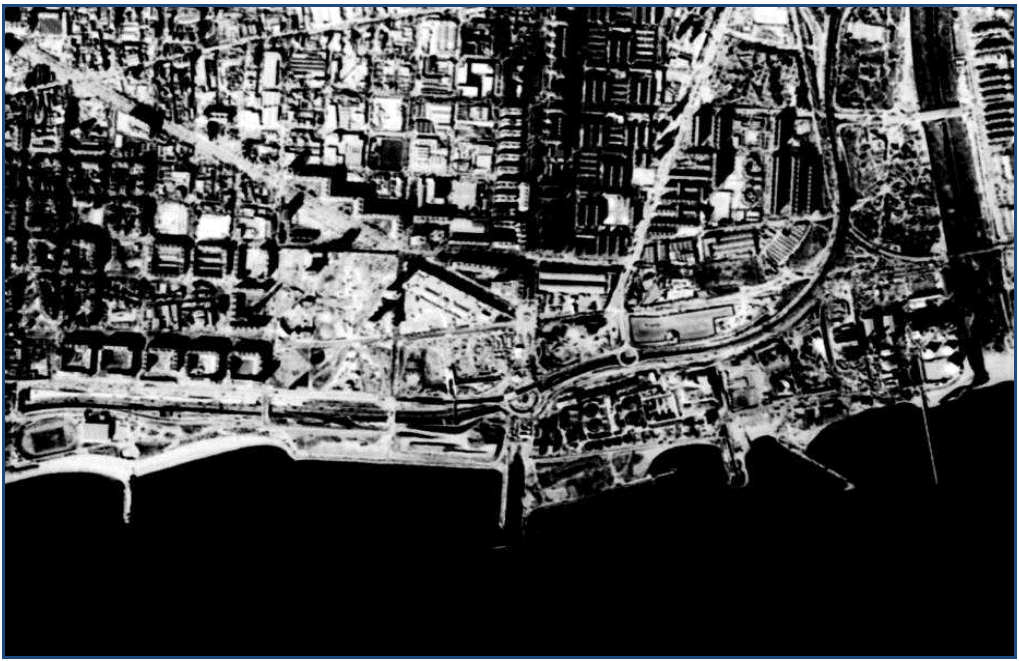

Figure n°6 : Le front de mer Besós-Diagonal et l'emplacement du futur Forum 2004. 1999, Photographie aérienne, au premier plan la station d'épuration. Un quartier en friche comme enjeu.

Un quartier en friche et l'effet d'accompagnement de ses infrastructures majeures pour la capitale catalane et leur entretien sur le long terme comme enjeu premier. Les Jeux olympiques de 1992 furent à part, car ils ont fait concilier trois éléments difficiles à coordonner et à juxtaposer : Identité politique + Identité économique + Identité urbanistique = cohésion sociale, économique et culturelle. Quant au Forum universel des cultures 2004 ; ces questionnements restent justement en suspend pour une partie de ses variables, car le temps n'a pas encore fait œuvre. Les habitants vont-ils s'approprier ou pas cette manifestation ou espace urbain novateur ?

Dernier point, concernant la France, celle-ci ne sait plus construire ni proposer des cahiers des charges convaincants en appui d'une diplomatie et influence mondiale en perte de vitesse. De plus avec le problème hypothécaire qui va se poser dans les années à venir, certains projets d'envergure comme une olympiade risquent d'être lettre morte pour ne plus devenir qu'un lointain souvenir. Notre dernière Exposition internationale remonte à 1937 et dernière olympiade à1924… Savons-nous toujours construire et faire des dépenses utiles et prospères dans le temps ? L'exemple de ces deux documents (1 et 2) qui accompagnent notre fin de raisonnement est là pour rappeler l'effet catalyseur et mobilisateur des forces vives d'une nation.

Figure n°7 : Communiqué de la présidence et du gouvernement de Catalogne sur les Jeux olympiques de Barcelone 1992 Original paru dans la presse catalane.

Fin août 1992, La Vanguardia.

**Traduction : Communiqué de la présidence et du gouvernement de Catalogne sur les Jeux olympiques de Barcelone 1992.**

Le gouvernement de la Generalitat exprime sa grande satisfaction et sa joie vis-à-vis de l'excellent déroulement des Jeux olympiques et félicite le COOB 92 en général et plus spécialement son président et Monsieur le Maire de Barcelone, Pasqual Maragall et Monsieur Josep Miquel Abad, conseiller délégué, pour la manière avec laquelle ils les ont organisés et il les remercie. Ce furent des Jeux et des jours inoubliables.
Il remercie également la contribution tout à fait décisive du président du COI, Monsieur Joan Antoni Samaranch.

Il exprime aussi sa reconnaissance à tous ceux, qui, d'une manière ou d'une autre y ont participé et les ont rendus possibles, plus particulièrement les sportifs qui ont eu une prestation remarquable et les volontaires. La générosité et l'esprit serviable des volontaires sont des motifs spéciaux de satisfaction parce qu'ils indiquent un état d'esprit très encourageant de notre jeunesse. Cette reconnaissance s'étend à toute la ville de Barcelone et à l'ensemble de la Catalogne qui, ont, non seulement apporté leurs efforts, mais aussi une grande joie et un formidable enthousiasme.

La Generalitat de Catalogne célèbre le fait que se soient pleinement réalisés les trois objectifs, qui, du point de vue du gouvernement de la Catalogne, devaient s'accomplir : améliorer substantiellement les infrastructures, organiser de bons Jeux olympiques qui rehaussent notre prestige face au monde entier et à travers cela, projeter partout, une image positive et attrayante de Barcelone et de la Catalogne. Dans ce même état d'esprit, le gouvernement de la Generalitat remercie les manifestations de catalinité pacifiques et de cohabitation que de nombreux citadins ont fait pendant les dernières semaines. Ils ont contribué très efficacement à faire que le monde entier sache que la Catalogne existe avec sa langue et sa culture propres, son drapeau et son hymne, sa capacité à exécuter des choses bien faites et sa volonté, en même temps, d'affirmation et de progrès.

Il célèbre également la bonne collaboration institutionnelle qu'il y a eu depuis la campagne pour obtenir la nomination de Barcelone jusqu'à la fin des Jeux. Cette collaboration institutionnelle qui s'est produite dans tous les domaines, a été spécialement évidente dans le COOB 92 ou ont travaillé ensemble la mairie de Barcelone, le gouvernement de la Generalitat, le gouvernement central et le Comité Olympique espagnol. Pour la Generalitat, cela a été un plaisir et un honneur de contribuer substantiellement au succès des Jeux. Je confie qu'avec le même effort, le succès des Jeux paralympiques va se répéter.
Finalement, le gouvernement de la Catalogne veut exprimer sa reconnaissance à sa Majesté le roi et à la famille royale pour sa présence et son appui.

En ce lendemain des Jeux, la Generalitat veut, en plus, exprimer sa confiance dans le futur. Confiance qui se base sur trois faits principaux. Le premier est la transformation positive qui s'est produite en Catalogne, pendant les années 1980. Le deuxième c'est un nouvel apport, qui, dans un sens, a représenté les Jeux et la bonne image qu'ils ont contribué à créer de la Catalogne. Le troisième c'est la conviction que nous devons avoir du fait que, si Barcelone et la Catalogne ont été capables de mener à bien un projet comme celui-là, cela veut dire que nous disposons de l'esprit d'initiative, de la capacité organisatrice et de l'esprit créateur qu'il est nécessaire d'avoir pour relever toutes les sortes de difficultés et pour que le pays progresse.
La Generalitat a confiance sur le fait que l'esprit patriotique, de dépassement et d'accomplissement, qui a présidé tout au long des Jeux, continuera à être présent. Le fait que cela continue ainsi, sera la meilleure garantie du progrès, de la stabilité et de l'identité de la Catalogne.

Barcelone, le 10 août 1992.

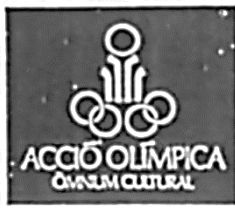

Figure n°8 : Communiqué de l'Action olympique sur les Jeux olympiques de Barcelone 1992 Original paru dans la presse catalane.

Fin août 1992, La Vanguardia.

**Traduction : Communiqué de l'Acció Olimpica (Publicité parue en août 99 dans l'ensemble de la presse locale).**

**Merci à Freedom for Catalonia**
Grâce à la devise Freedom for Catalonia, des millions de personnes ont découvert qu'il y a une nation qui s'appelle la Catalogne qui n'a pas résolu sa question nationale, sa capitale ayant été le siège des Jeux. Nous nous félicitons donc, non seulement pour le bon accueil que cette devise a obtenu en Catalogne, mais aussi pour celui qu'elle a eu dans la presse internationale.

**Merci aux drapeaux**
Grâce aux drapeaux catalans, que des milliers de citadins ont exhibés lors du passage du flambeau sur les balcons et dans les stades, le monde entier connaît la personnalité de la nation catalane : une nation millénaire et consolidée qui veut continuer à travailler pour la construction européenne, le progrès et la modernité. Une personnalité qui nous pousse à construire le futur avec des aspirations de liberté, de force, de responsabilité, de l'amour pour la paix et avec la ferme volonté de maintenir l'identité linguistique et culturelle propre.

**Merci aux personnes**
Merci aux milliers de personnes, qui, dans toute la Catalogne, ont contribué à un quelconque aspect de la campagne, qui ont investi temps et efforts pour préparer la plus importante projection de la Catalogne de l'histoire ; nous avons corrigé l'image stéréotypée et erronée qui aurait pu être donnée sur notre pays. Le danger que les Jeux se soient fait à Barcelone, et que personne ne soit arrivé à savoir qu'ils se faisaient en Catalogne, s'est évité. Nous y sommes parvenus.

**Merci aux Jeux**
Grâce aux Jeux olympiques de Barcelone, la Catalogne a fait un pas important dans tous les domaines. On ne peut pas en douter.

**Pour tout cela**
Il est clair que le succès des Jeux et la projection nationale de la Catalogne sont des concepts complémentaires et que les rendre plus puissants, sans que cela constitue un problème, a fait considérablement augmenter la participation et l'enthousiasme populaire. Ces Jeux ont été brillants. La Catalogne a atteint ce qu'elle prétendait et nous avons démontré au monde entier qu'elle est un pays dans lequel on peut avoir confiance. .

Il nous reste beaucoup de défis à relever. Il y a beaucoup de tâches à réaliser: Avec optimisme et efficacité. Jour après jour dans tous les domaines, il faut continuer à travailler.

Barcelone, le 10 août 1992.

## Conclusion

Il faut rester prudent sur le devenir de ce nouvel espace public, l'espace – port Forum à Barcelone. Les Barcelonais et les investisseurs décideront de sa destinée économique, touristique et sociétale. On peut toutefois souligner que Barcelone reste une ville à part à l'échelle mondiale dans le domaine de l'expérimentation urbanistique à travers l'organisation et l'évaluation de l'événementiel sportif et culturel. La capitale catalane joue sur les grands événements pour produire une trame urbaine et une croissance économique pour tous, ceci dans le cadre de proposition pour réaménager la ville tout en accompagnant la globalisation de son économie. L'intérêt économique d'organiser une manifestation géante éphémère est toujours d'actualité et se vérifiera dans le futur avec le rôle des pays asiatiques au sein du CIO et du BIE (Bureau international des expositions). L'important est de développer une argumentation rationnelle pour communiquer sur le bien-fondé économique de la manifestation. Le développement durable et la Responsabilité sociale des entreprises (RSE) sont des éléments prédominants pour des événements éphémères qui se veulent durables dans le temps pour les générations futures. Il faut tenir compte des principales dimensions des impacts de l'événement et de la stratégie planificatrice en amont-aval, tenir compte des temporalités de l'événementiel pour son évaluation, tenir compte des grands événements déjà produits dans le pays organisateur, tenir compte du contexte politico-économique (pays en récession au moment de l'événement, pays en croissance faible, en croissance moyenne ou forte, voire reconstruction d'un pays ou d'une ville après la guerre, catastrophe, crise économique), tenir compte du régime institutionnel organisant l'événement, dictature, démocratie autoritaire ou régime démocratique parlementaire – présidentiel, tenir compte des experts et différents spécialistes de la question BIE, CIO, bureau d'audit, universitaires, grandes entreprises partenaires et les collectivités locales associées, tenir compte des objectifs communs ou divergents des différents acteurs et tenir compte comme pour les dossiers de candidatures actuels des grandes réussites passées comme en Espagne avec Barcelone 1992 Jeux olympiques ou Séville 1992 exposition universelle sur le plan touristique.

## Bibliographie

# ANNEXES

<u>Expérience d'autres villes-hôtes</u>

Les villes se servent de plus en plus de grandes manifestations, comme les Jeux Olympiques, pour élargir leurs perspectives économiques à court et à long terme et accroître leur visibilité sur la scène internationale. Elles espèrent ainsi attirer des touristes, obtenir une couverture médiatique à l'échelle mondiale, stimuler rapidement le renouvellement de leurs infrastructures et favoriser la construction d'installations sportives et de loisirs.

Les Olympiques sont généralement vues comme l'occasion toute indiquée pour mettre à exécution des projets immobiliers et des plans d'aménagement à long terme, créer des débouchés pour le tourisme et les investisseurs et aider les villes-hôtes à réaliser leurs objectifs économiques. L'expérience des villes-hôtes indique que la tenue des Olympiques a certes donné lieu à maintes *occasions* à des développements et des changements.

<u>Dans ce contexte, les Jeux peuvent :</u>

1. Servir de catalyseur pour améliorer les infrastructures et réaliser des projets d'aménagement d'envergure.
2. Stimuler le tourisme avant l'événement, de sorte que le niveau d'activité culmine pendant l'année des Jeux, puis demeure plus élevé que la moyenne au cours des années ultérieures.
3. Créer des emplois à court terme pour la prestation de services et la préparation du grand événement, des activités qui se traduiront ultérieurement par un bilan migratoire positif pour la région hôte.
4. Inciter les gens à visiter une ou plusieurs fois la région ou à venir y investir.
5. Favoriser l'élaboration de nouveaux quartiers et la régénération urbaine.
6. Fournir l'occasion de construire des logements abordables pour les résidents à faibles revenus.

<u>En revanche, les villes-hôtes émettent aussi un certain nombre de sérieuses mises en garde et réserves, dont les suivantes :</u>

1. Le défi d'équilibrer les attentes venant de l'extérieur (échelle internationale), tout en atteignant les objectifs fixés à l'échelle locale.
2. Les utilisations spécialisées des installations créées pour les Jeux d'*hiver*.
3. Le coût des infrastructures, comme le transport, l'hébergement touristique, les complexes pour athlètes et les centres de média. Bien que ces éléments constituent un legs considérable, ils peuvent aussi représenter un risque financier accru.
4. Les changements à long terme observés dans le tourisme et le secteur de l'immobilier ne résultent pas entièrement de la tenue d'un grand événement, car d'autres facteurs interviennent, tels que la migration, l'emploi et les niveaux de revenus. Si le taux d'occupation des hôtels dépasse la moyenne durant la période menant à la tenue des Olympiques, ce secteur est souvent caractérisé par une offre excédentaire pendant quelque temps après l'événement.
5. L'accroissement du nombre de nouveaux touristes et de touristes internationaux est fonction d'un programme de marketing dynamique.
6. Le nombre de visiteurs venus pour les Jeux et les coûts connexes peuvent « gêner » les visiteurs réguliers. De plus, l'effet de substitution interne peut faire en sorte que l'événement attire des touristes au détriment d'autres centres de villégiature situés dans les régions avoisinantes.
7. L'intensification de la demande de logements locatifs à court terme découle surtout du nombre de travailleurs qui participent à la préparation des Jeux plutôt que des spectateurs. La conversion de logements locatifs à prix modique peut entraîner le déplacement de locataires.

2006, Société canadienne d'hypothèques et de logement
Imprimé au Canada
Réalisation : SCHL 02-11-06

Résumé concernant les impacts d'une olympiade par la société canadienne d'Hypothèque et de logement. 2006

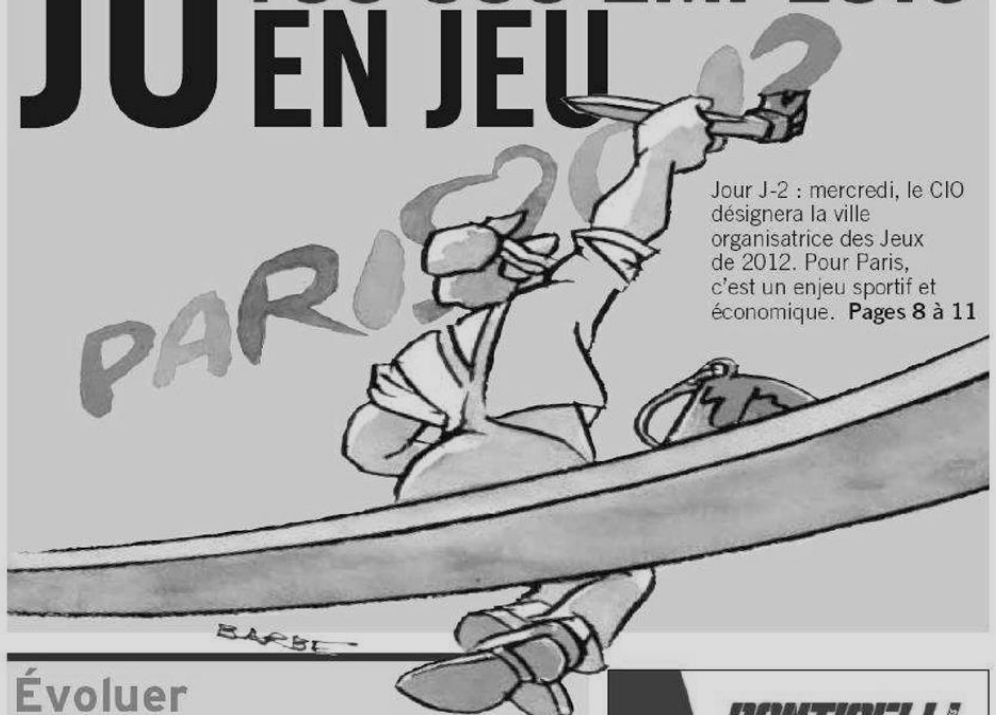

Le principe d'espérance de Jeux olympiques de 2012. 100 000 emplois en jeu.

(Figaro, Juillet 2005)

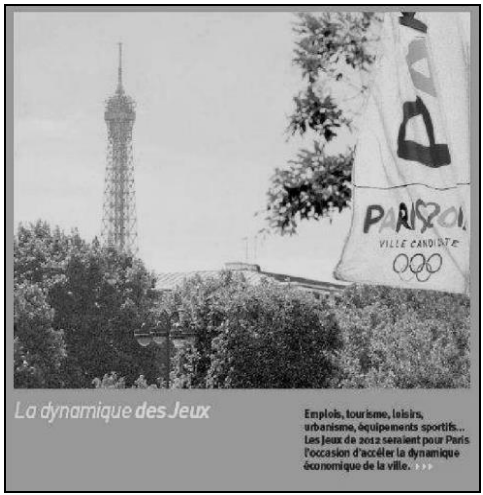

Le degré zéro des contraintes et évacuations de tout risque d'endettement budgétaire et d'aléas négatifs.
Paris Service 02/04/2005

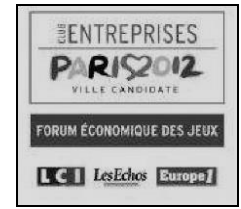

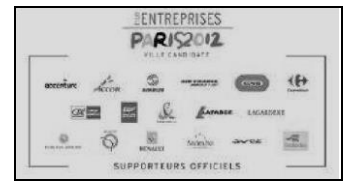

L'appel des institutionnels est toujours de rigueur, le 4/02/2005 colloque du club des entreprises où jamais les risques de telle manifestation ne sont invoqués. Évaluation sans recul critique.

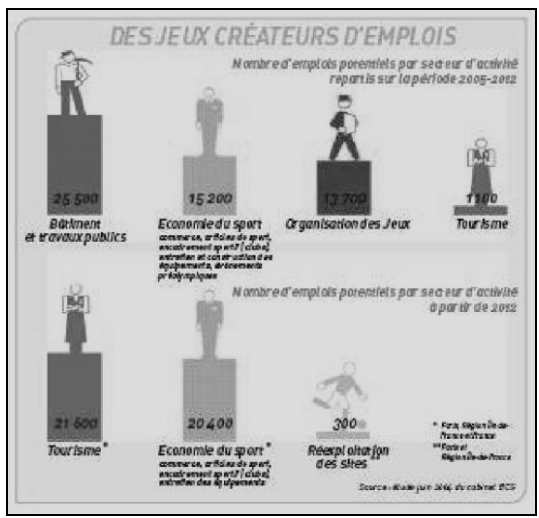

Le gain d'emploi est toujours mis en avant avec parfois des contre-exemples ou substitution d'emploi en cas de non tenue.

Paris Service 02/04/2005

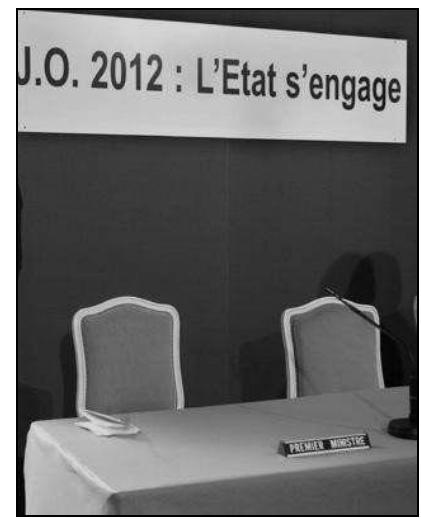

Un État s'engage, car le risque de déficit budgétaire est bien là, voire inhérent à la tenue de Jeux. Les bénéfices des droits TV / billetterie / sponsoring sont toujours en deçà des sommes que l'État engage pour la construction des infrastructures ou la sécurité des sites.
Paris, Matignon, 06/2005

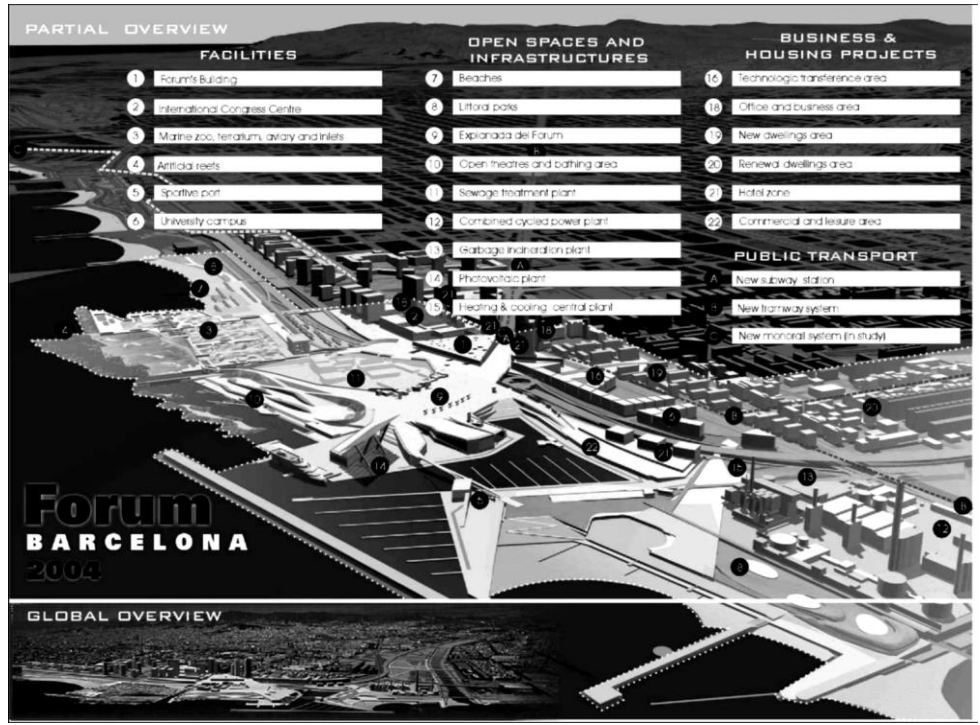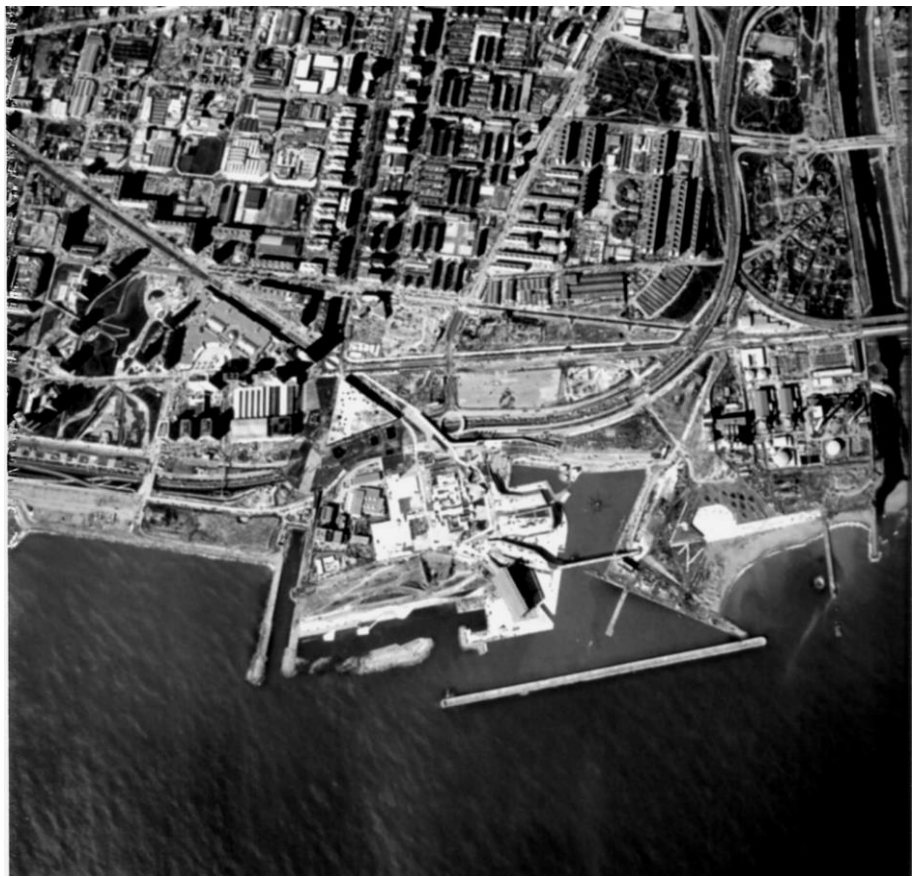

Barcelone, du Forum universel des cultures, d'un événementiel culturel, exposition internationale de l'UNESCO à la création d'un véritable quartier de ville portuaire, touristique (affaires) et immobiliers de luxe, zone universitaire et de sièges sociaux. 2002/2008

Les grands événements à caractère touristique, qu'ils soient artistiques (Festival International du Film de Cannes), culturels (Exposition Universelle), festifs (Carnaval de Rio), sportifs (Jeux Olympiques) etc., ont pris de nos jours une telle ampleur qu'ils ont nécessairement un impact sur l'économie locale, régionale, voire nationale. La connaissance de leur incidence réelle est devenue indispensable pour mieux conduire les politiques de développement et de tourisme.

Le présent ouvrage tente de dégager des méthodes d'évaluation visant à mesurer de la manière la plus pertinente possible les retombées des événements touristiques.

Après un propos préliminaire consacré à la place de l'événementiel touristique dans la littérature académique, les contributions des auteurs, regroupées autour de cinq parties, cherchent à répondre aux questions suivantes : Les méthodes traditionnelles d'évaluation sont-elles incontournables ? De la diversité des événements à une mesure généralisable ? Peut-on créer ou pérenniser une identité à partir d'un événement touristique ? Peut-on innover en matière d'évaluation ? L'éphémère touristique est-il durable ?

Les réponses à ces interrogations s'appuient sur des études réalisées dans douze pays différents par des universitaires et des professionnels du tourisme. Leur coordination a été réalisée au sein du groupe de recherche « Tourisme : Marchés et Politiques » qui s'est constitué, il y a près de dix ans, au sein de l'Université de Nice Sophia-Antipolis, à l'initiative de Jacques Spindler. C'est cette équipe qui a publié en 2004, dans la même collection, le *Tourisme au XXIe siècle*, et qui a organisé, en 2005, dans le cadre d'une convention de recherche avec la Direction du tourisme, un colloque international sur le thème de l'évaluation des événements touristiques.

Que cet ouvrage puisse contribuer à l'exploration de nouvelles pistes de recherche afin d'enrichir les méthodes traditionnelles d'évaluation.

*Jacques* **SPINDLER** *est Professeur en Sciences de gestion à l'Université de Nice Sophia-Antipolis où il dirige l'Institut d'Administration des Entreprises. David* **HURON** *est Maître de Conférences en Sciences de gestion à l'Université de Nice Sophia-Antipolis.*

Illustration : © Michèle BONREPAUX, *Holidays*, 2009

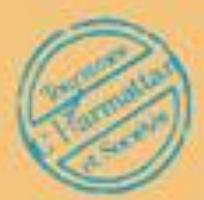
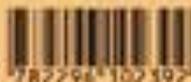

Collection
Tourismes et Sociétés
dirigée par
Franck Michel

ISBN : 978-2-296-10239-2

45 €